\documentclass[%
 reprint,
nofootinbib,
 amsmath,amssymb,
 aps,
prd,
]{revtex4-2}

\usepackage[hidelinks,colorlinks=true,linkcolor=blue,citecolor=blue]{hyperref}
\newcommand\aap{A\&A}                
\newcommand\aj{AJ}                   

\newcommand\apjl{ApJ}                

\newcommand\mnras{MNRAS}             
\newcommand\planss{Planet. Space~Sci.} 
\usepackage{natbib}
\usepackage{CJK}
\usepackage{tabu}
\usepackage{multirow}
\usepackage{booktabs}
\usepackage{graphicx}
\usepackage{dcolumn}
\usepackage{bm}

\begin{document}


\title{Amplified capture rate of dark matter in compact binaries and constraints on bosonic dark matter from GW170817}

\author{János Takátsy$^{1,2}$}
\email{janos.takatsy@nbi.ku.dk}
\author{Lorenz Zwick$^{1,2}$}
\author{David O'Neill$^{1}$}
\author{Johan Samsing$^{1,2}$}
\affiliation{%
$^{1}$Niels Bohr International Academy, The Niels Bohr Institute, Blegdamsvej 17, DK-2100, Copenhagen, Denmark\\
$^{2}$Center of Gravity, Niels Bohr Institute, Blegdamsvej 17, 2100 Copenhagen, Denmark.}%

\date{\today}

\begin{abstract}
The accretion of dark matter (DM) onto compact objects and the potential gravitational collapse of neutron stars due to this accretion has become a promising indirect probe of DM properties, complementing terrestrial experiments. We show that the accretion flux of DM on stellar objects is amplified in binary systems due to the complex gravitational interaction of said particles with the binary. We perform few-body Monte Carlo simulations to show that this amplification factor is $\sim4-5$ for circular binaries, small DM velocity dispersions and mass ratios $q\gtrsim0.3$. We use this factor to improve previous constraints on the scattering cross section of non-annihilating bosonic DM with baryonic matter, and derive upper bounds on this cross section from the observation of the binary NS merger associated with GW170817. We also show that the maximally accretable mass fraction of DM by binary NSs is $\lesssim10^{-3}$, even for extreme DM densities only possible in DM spikes, due to the dynamical friction exerted by the ambient DM.
\end{abstract}

\maketitle


\section{Introduction}
\label{sec:introduction}

The accretion of dark matter (DM) onto stars and compact objects such as neutron stars (NSs) has emerged as a promising indirect probe of DM properties, especially in the regime where DM particles interact weakly but sufficiently with baryonic matter to be gravitationally captured and subsequently thermalized within the star’s interior \cite{Press1985,Gould:1987ww,Gould:1987ir,Jungman:1995df,Kumar:2012uh,Kappl:2011kz,Busoni:2013kaa,Bramante:2017xlb,Leane:2024bvh,Bramante:2023djs,Singh:2022wvw,Khadkikar:2025awf,Robles:2025dlv}. Crucially, the rate at which DM particles are captured by stars is governed by their scattering cross section with stellar constituents, such as nucleons or leptons. This makes them valuable astrophysical tools complement to terrestrial direct detection experiments, which probe the same underlying interactions in a different kinematic regime \cite{XENON:2023cxc,XENON:2024znc,PandaX:2022osq,LZ:2022lsv,LZ:2024zvo}.

The influence of DM on stellar objects provide a wide range of potentially observable phenomena. The annihilation of DM particles inside the Sun could manifest as observable radiation of neutrinos \cite{Super-Kamiokande:2011wjy,Super-Kamiokande:2015xms,ANTARES:2016obx,ANTARES:2016xuh,IceCube:2016dgk} or other particles \cite{Batell:2009zp,Schuster:2009au,Bell:2011sn,Feng:2016ijc,Leane:2017vag}. The energy transport inside the Sun might also be altered by the presence of DM \cite{1990ApJ...352..654G,1990ApJ...352..669G,Vincent:2013lua,Geytenbeek:2016nfg}. The impact of DM on compact objects also spans a variety of physical effects. Dynamical drag from ambient DM may alter binary orbital evolution \cite{Eda:2013gg,Eda:2014kra,Hannuksela:2018izj,Hannuksela:2019vip,Kavanagh:2020cfn,Annulli:2020lyc,Traykova:2021dua,Coogan:2021uqv,Vicente:2022ivh,Cole:2022ucw,Speeney:2022ryg}, while the accumulation of DM can modify the mass-radius relation of NSs, potentially affecting their stability and observable properties \cite{Shakeri:2022dwg,Rutherford:2022xeb,Giangrandi:2022wht,Shirke:2023ktu,Thakur:2023aqm,Ivanytskyi:2019wxd}. Additionally, DM annihilation within NSs and kinetic heating through scatterings can modify the cooling evolution of NSs, providing observable signatures that can be used to infer DM properties \cite{Kouvaris:2007ay,Busoni:2021zoe,Raj:2017wrv,deLavallaz:2010wp,Baryakhtar:2017dbj,Bell:2018pkk,Camargo:2019wou,2010A&A...522A..16G}. Gravitational wave (GW) observations, particularly measurements of tidal deformability during NS mergers, offer another avenue to detect the presence of DM within NSs \cite{Karkevandi:2021ygv,Koehn:2024gal,Das:2021wku,Ellis:2018bkr,Nelson:2018xtr}. Furthermore, numerical relativity simulations of NS mergers that include DM components have shown that DM accumulated inside NSs can influence the dynamics and outcomes of such events \cite{Bauswein:2020kor,Ruter:2023uzc,Giangrandi:2025rko,Emma:2022xjs}. In extreme cases, for non-annihilating DM -- especially bosonic candidates capable of forming a Bose--Einstein condensate -- accretion beyond a critical threshold can trigger gravitational collapse within the NS, resulting in BH formation \cite{Goldman:1989nd,Kouvaris:2010jy,Kouvaris:2011fi,McDermott:2011jp,Guver:2012ba,Bell:2013xk,Bramante:2013nma,Bertoni:2013bsa,Garani:2018kkd,Liang:2023nvo}. Consequently, the observation of NSs in merging binaries -- or more precisely, the lack of observations of collapsed NSs --, particularly in old elliptical galaxies where DM densities are expected to be high, places strong upper bounds on the DM--nucleon interaction cross section across a range of DM particle masses \cite{Singh:2022wvw,Bramante:2017ulk,Khadkikar:2025awf}. Electromagnetic counterparts, such as a kilonova \cite{Cowperthwaite:2017dyu,Villar:2017wcc,LIGOScientific:2017ync}, together with the GW signal of a merger of two compact objects in the mass range $\sim1.2-2.2~M_\odot$ expected for NSs, would unambiguously tell that at least one of the binary components was a NS. NS--BH mergers with a low-mass BH component of $\sim5~M_\odot$ are also expected to produce an electromagnetic counterpart, depending on the NS equation of state and the BH spin \cite{Barbieri:2019kli}. Another indication of a NS component would be the observation of a GW phase shift due to the tidal deformation of the NS \cite{Singh:2022wvw,Khadkikar:2025awf}, however, this dephasing is also expected to be much lower for NS--BH binaries with a massive BH component.

The gravitational capture of weakly interacting particles by astrophysical bodies has been studied for decades, beginning with foundational work on DM capture by stars and white dwarves \cite{Press1985,Gould:1987ww,Gould:1987ir,Jungman:1995df,Kumar:2012uh,Kappl:2011kz,Busoni:2013kaa,Bramante:2017xlb,Bell:2020jou,Bell:2020lmm,Bell:2020obw,Anzuini:2021lnv}. In the context of NSs, their high density and deep gravitational wells make them particularly efficient at capturing DM particles. These studies traditionally focus on accretion in isolated stars or compact objects. However, a significant fraction of stars, including those that evolve into NSs, form and spend most of their lifetime in binary systems \cite{2012Sci...337..444S,Postnov:2014tza,2024ApJ...977..203C}. Moreover, the progenitors of binary NS mergers are expected to have GW inspiral timescales longer than their stellar evolution timescales. For the progenitor of GW170817, an inspiral time of $t_\mathrm{insp}\gtrsim3-6.8$~Gyr was estimated based on the properties of the host galaxy, while the stellar evolution timescale for a $20~M_\odot$ star is $t_\mathrm{st}\gtrsim100$~Myr \cite{Pan:2017jem,Margutti:2017cjl,Blanchard:2017csd,Belczynski:2017mqx}. In such systems, due to the complex orbital motion of DM particles, the gravitational focusing and the probability of capture of these particles can increase due to repeated close encounters of the DM particles with the NSs, speeding up the accumulation of DM inside these objects. This amplification of the effective cross section for DM accretion has been noted in the analogous context of stellar hardening of supermassive BH binaries and the amplified rate of tidal disruption events near such binaries \cite[e.g.][]{Quinlan:1996vp,Coughlin:2016wyp,Melchor:2024}, but its implications for DM capture in binary stars has not yet been quantified.

In addition, the presence of DM in binary NS systems can influence the inspiral time through dynamical friction. The extracted orbital energy accelerates binary hardening and thus can significantly shorten the time it takes the binary to merge. This, however, will impose an upper bound on the total amount of dark matter that NSs can accrete before the merger of the binary, even in regions with exceptionally high dark matter densities, such as those suggested in DM spikes \cite{Gondolo:1999ef,Merritt:2002vj,Ullio:2001fb,Gnedin:2003rj,Sadeghian:2013laa}.

In this study, we quantify the enhancement of DM accretion in compact binaries using Monte Carlo simulations of test-particle scattering in binary gravitational potentials. We calculate the amplified capture rate and derive updated constraints on the cross section of bosonic DM based on the possible collapse of the NSs into BHs and the lack of observation of these binary systems. We also provide upper bounds on the DM fraction inside binary NSs at merger, based on the effects of dynamical drag on binary evolution. Our results improve upon previous constraints obtained from isolated NSs.

This paper is structured as follows...

\section{Particle flux in binaries}

Assuming a single static object with mass $m$ embedded in a cloud of non-interacting particles with a number density $n$ and a uniform velocity $\boldsymbol{v}$ far away from the object, the flux of particles crossing a spherical surface with radius $r$ is
\begin{equation}
    \mathcal{F}_s(v) = n \Sigma(v) v \: ,
\end{equation}
where $v=|\boldsymbol{v}|$ and $\Sigma(v)$ is the cross section that also includes gravitational focusing:
\begin{equation}
    \Sigma_s(v) = \pi b_\mathrm{max}^2 = \pi r^2 \left( 1+ \frac{2G m}{v^2 r} \right) \: ,
    \label{eq:GrCrSec}
\end{equation}
where we used that the impact parameter $b$ of a test-particle orbit with a pericenter distance $r_\mathrm{p}$ is
\begin{equation}
    b^2 = r_\mathrm{p}^2 \left( 1+ \frac{2G m}{v^2 r_\mathrm{p}} \right) \: .
    \label{eq:rpb}
\end{equation}
For an isotropic velocity distribution $f(v)$ the total flux becomes:
\begin{equation}
    \left\langle\mathcal{F}_s\right\rangle = n \int \mathrm{d}v \, 4\pi v^3f(v) \Sigma_s(v) \equiv n \left\langle\Sigma_s\right\rangle v_\chi \: .
    \label{eq:Fave}
\end{equation}
We can assume that $f(v)$ is a Maxwell-Boltzmann distribution with a 3D velocity dispersion $v_\chi$:
\begin{equation}
    f(v) = \left( \frac{3}{2\pi v_\chi^2} \right)^{3/2} \exp\left(-\frac{3v^2}{2v_\chi^2}\right) \: .
\end{equation}
To evaluate the integral given in Eq.~\ref{eq:Fave} we first consider the limit of small radii, in which the escape velocity is much larger than the velocity dispersion, i.e. $v_\mathrm{esc}=\sqrt{2Gm/r}\gg v_\chi$. We obtain an estimate for the average cross section as
\begin{equation}
    \left\langle\Sigma_s\right\rangle \approx \sqrt{6\pi} \frac{2Gm}{v_\chi^2} r \: .
\end{equation}

\begin{figure}[!t]
    \centering
    \includegraphics[width=0.45\textwidth]{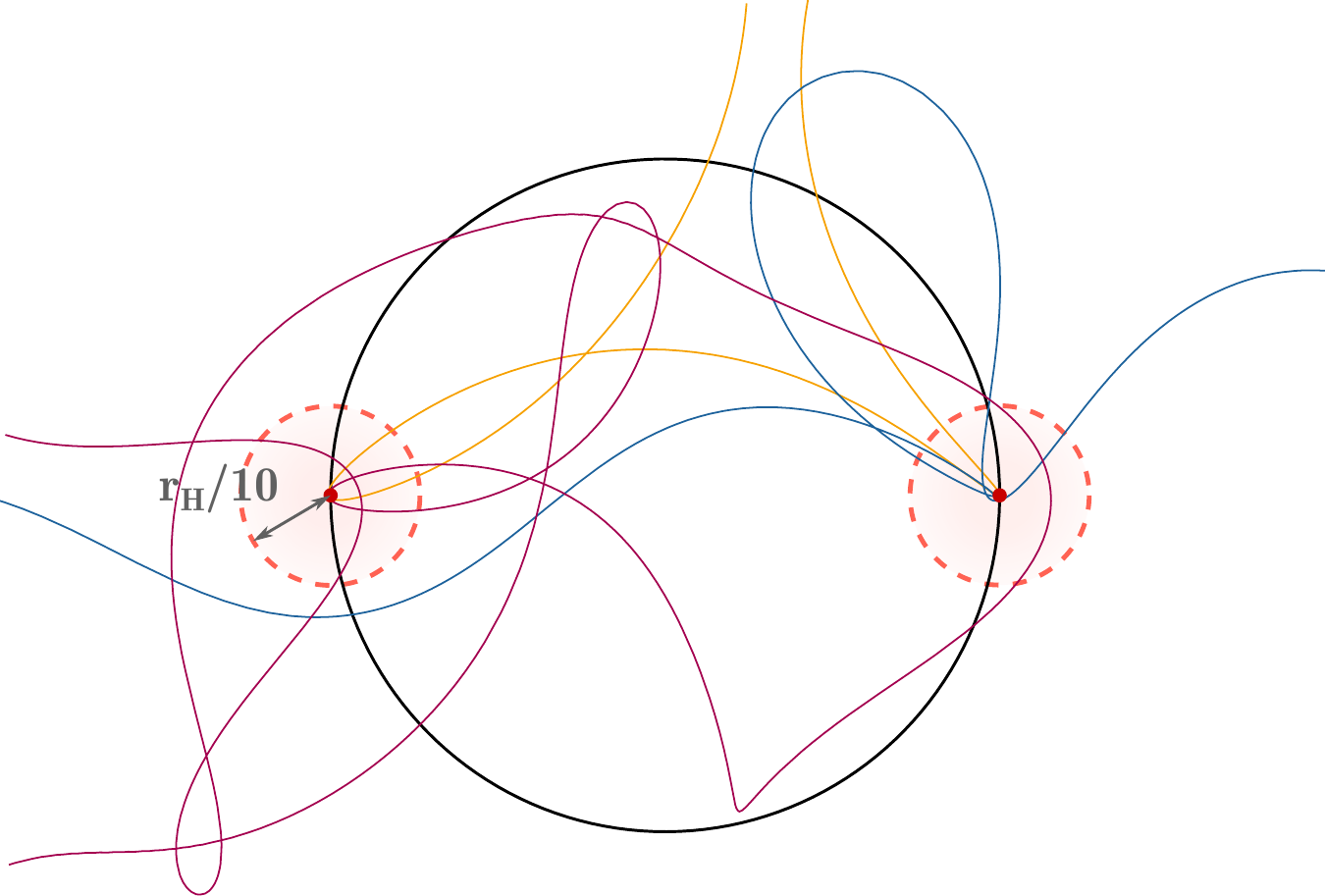}\\
    \vspace{0.4cm}
    \includegraphics[width=0.45\textwidth]{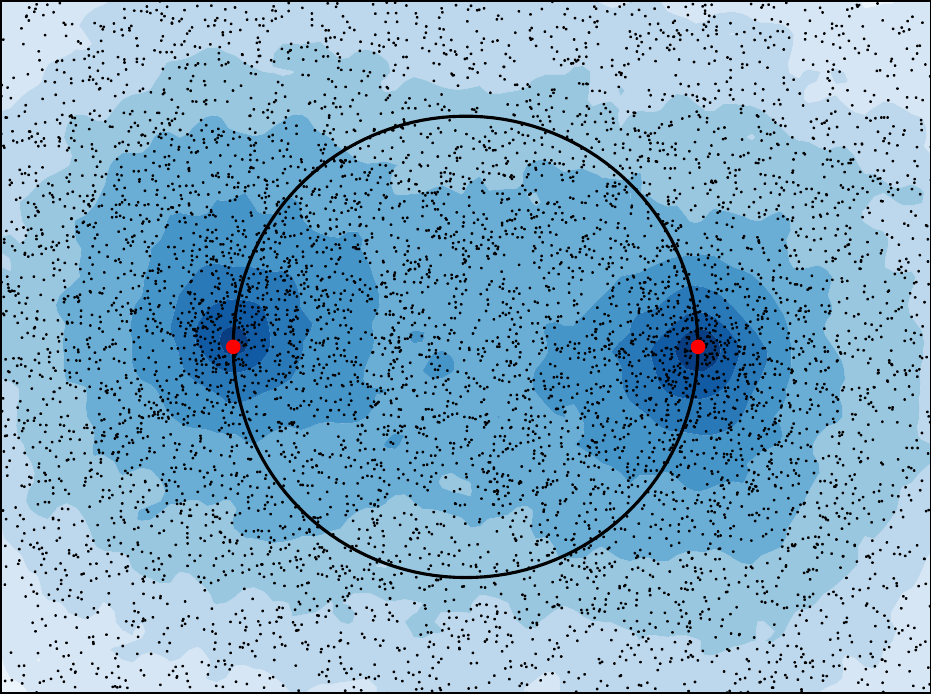}
    \caption{\textit{Top:} Illustration of a few test-particle orbits in the corotating frame of a binary NS system. If the particle velocity at infinity is low compared to the binary orbital velocity, its orbit can be efficiently redirected and thus many particles can have multiple encounters with one (blue) or with both (yellow and purple) objects. The radii of the close vicinity of compact objects have been enlarged for visibility. \textit{Bottom:} The density profile of DM in the plane of the binary rotating counter-clockwise at a given time instance. The center of the DM profiles is lagging behind the NSs, producing an overall drag force. The density profile was produced using a limited number of test-particle simulations.}
    \label{fig:illustration}
\end{figure}
The flux of non-interacting particles on objects inside binary systems cannot be calculated analytically, as the test-particle orbits become non-integrable. As a result, these particles can have multiple close encounters with the binary components. Fig.~\ref{fig:illustration} illustrates this by showing some selected particle orbits that enter the close vicinity of either of the binary components multiple times. Therefore the total flux of particles on one of the binary components can increase. Then the cross section for non-interacting particles crossing the surface of an object inside a binary will become:
\begin{equation}
    \Sigma_b(v) = R_q\left( v/V_b \right) \Sigma_s (v) \: ,
\end{equation}
where $R_q$ is an amplification factor that depends on the mass ratio $q=m_2/m_1$, and $v/V_b$, where $V_b \equiv \sqrt{GM/a}$ is the relative velocity of circular binary components, with $M=m_1+m_2$ being the total mass and $a$ the semi-major axis of the binary. Due to the non-integrability of test-particle orbits, this amplification factor can only be calculated numerically. Note that in principle $R_q$ could be a function of $r/a$ as well, however, we find from our simulations that for a fixed $v$, $R_q$ is constant for multiple orders of magnitude in $r$ when $r/a\ll1$. Moreover, due to the scale-free property of gravitationally interacting systems, $R_q$ cannot be a function of $v$ and $V_b$ individually, but only of $v/V_b$. Taking this into account the total flux of particles on one of the binary components becomes:
\begin{align}
    \left\langle\mathcal{F}_b\right\rangle &= n \int \mathrm{d}v \, 4\pi v^3f(v) R_q(v/V_b) \Sigma_s(v) \nonumber \\
    &\equiv n \left\langle\Sigma_b\right\rangle v_\chi \equiv n \left\langle R_q\right\rangle \left\langle\Sigma_s\right\rangle v_\chi \: ,
\end{align}
where we have introduced the averaged amplification factor:
\begin{equation}
    \left\langle R_q\right\rangle \left(v_\chi/V_b\right) \equiv \frac{\left\langle\Sigma_b\right\rangle}{\left\langle\Sigma_s\right\rangle} \: .
\end{equation}

\subsection{Simulation setup and results}

The total flux $\mathcal{F}_b$ and thus the amplification factor $R_q$ can only be determined through numerical integration of the restricted three-body problem. We create a Monte Carlo setup similar to that of Ref.~\cite{Quinlan:1996vp} used to determine the hardening rate of massive black hole binaries due to dynamical drag from a dense stellar environment. We fix the binary with component masses $m_1$ and $m_2$ to a circular orbit with semi-major axis $a$. The velocity of the test-particle at infinity is also fixed to $v$. The three-body scattering is then characterized by four independent parameters, the impact parameter $b$ to the center of mass of the binary, and three parameters characterizing the binary orbit, the inclination $i$, the longitude of the ascending node $\Omega$ and the mean anomaly $l$ at some reference time. These are picked from four uniform distributions, one for the square of the impact parameter in a range $[0,b_\mathrm{max}^2]$, one for the cosine of the inclination between $[-1,1]$ and two for the longitude of the ascending node and the mean anomaly in the range $[0,2\pi]$. As particles with a large impact parameter do not cross the trajectory of the binary objects, we truncate the distribution of impact parameters at a value corresponding to a scaled pericenter distance $r_\mathrm{p}^{\mathrm{max}}/a = 4$ according to Eq.~\eqref{eq:rpb}.


The coordinate system is fixed so that the binary center of mass is at the origin and the test particles have position and velocity vectors at infinity, $(x,y,z)=(b,0,\infty)$ and $(v_x,v_y,v_z)=(0,0,-v)$. Since at large distances the test-particles only see the binary as a single central object with a mass $M=m_1+m_2$, the particles are analytically propagated along a Keplerian orbit about a point mass $M$ to a radius $r=50 a$. Then the motion of the test-particle is integrated with a fourth-order Runge-Kutta integrator with adaptive step size and the fractional error per step set to $10^{-9}$. The forces from the binary objects are not softened. All the orbits are unstable and thus eventually become expelled except for a set of initial condition with a volume of zero in the initial parameter space. Thus we terminate the integration if the test-particle crosses the sphere of radius $r=50 a$ with a positive total energy. We also terminate the integration if they last more than $10^6$ steps. We record close encounters with one of the binary components if the approach is within a separation $r<0.1 r_H$, where $r_H = a (m_i/3M)^{1/3}$ is the radius of the Hill sphere around the $i$th object. Then, the velocity of the test particle is transformed into the co-moving frame of that object and its closest approach is calculated according to a Keplerian trajectory. This number is then recorded. The flux of particles at a given radius is then calculated using the cumulative number of encounters with a closest approach smaller than that radius.

We split the range of impact parameters to five intervals corresponding to scaled pericenter distances $r_\mathrm{p}/a$ of $[0,0.25]$ $[0.25,0.5]$, $[0.5,1]$, $[1,2]$ and $[2,4]$. We first run a simple Monte Carlo simulation until a certain number of close encounters is recorded. Then we run the simulation again for a larger number of initial conditions with an importance sampling according to the number of close encounters recorded in each interval. This process reduces the statistical error of the recorded flux for a given number of initial conditions.

As mentioned earlier we find that the amplification factor $R_q = \mathcal{F}_b/\mathcal{F}_s = \Sigma_b/\Sigma_s$ is constant for multiple orders of magnitude for radii smaller than $0.1 r_H$. For $v\gg V_b$ this amplification factor should go to $1$ as in this limit, the binary resembles two stationary point particles, each with mass $M/2$ to a test particle with fixed velocity $v$.

\begin{figure}[!t]
    \centering
    \includegraphics[width=0.49\textwidth]{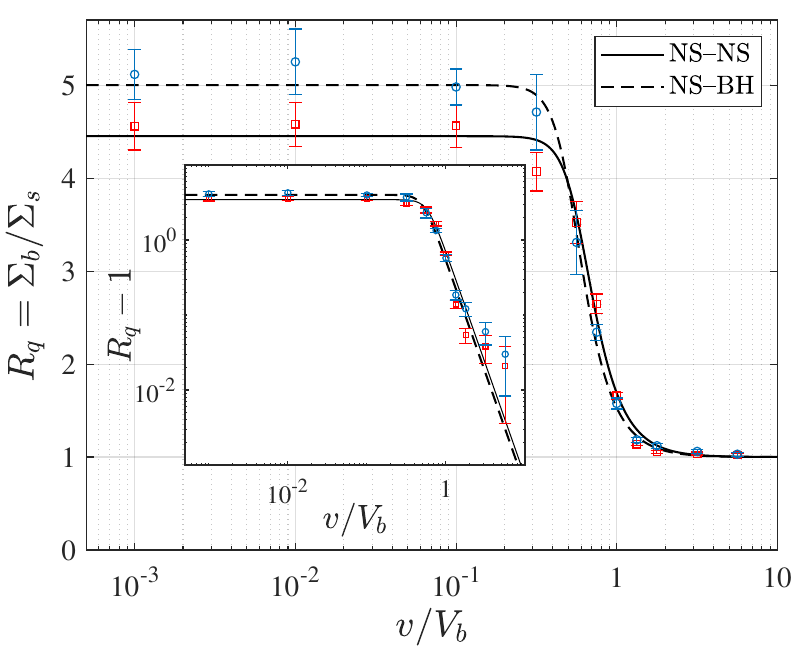}
    \caption{The amplification factor of the cross section for DM accretion in binary systems for a homogeneous and isotropic DM background as a function of DM velocity over binary orbital velocity. Red data points and solid fitted line correspond to a binary NS with $q=1$, while blue data points and the dashed line correspond to a NS--BH binary with $q=1.4/5=0.28$.}
    \label{fig:Rv}
\end{figure}

We run our simulations for two different mass ratios, $q=1$ and $q=0.28$, which correspond to two equal-mass NSs and a NS--BH system with masses of $1.4~M_\odot$ and $5~M_\odot$, respectively. We chose the mass of the BH so that an electromagnetic counterpart is still expected to be produced by the merger of the binary, which would unambiguously tell that the lower-mass object has not yet collapsed into a BH due to DM accretion. The results for $R_q(v/V_b)$ are shown in Fig.~\ref{fig:Rv}. For low velocities $v$ the amplification factor saturates at $\sim4.5$, while it drops sharply for $v/V_b\gtrsim1$. We fit the results with the empirical formula:
\begin{equation}
\label{eq:fit}
    R(x) = 1 + \frac{R_0}{\left[1+(x/w)^6\right]^{1/2}} \: ,
\end{equation}
where $x\equiv v/V_b$. The fitted parameters can be found in Table~\ref{tab:params}. We find that the formula above gives a good fit for the simulated data, although the data points for large $v/V_b$ seem to deviate slightly from the $x^{-3}$ power-law dependence.

\renewcommand{\arraystretch}{1.5}
\begin{table}[]
\begin{tabular}{ccc}
\toprule
\quad $q=m_2/m_1$ \quad & \quad $R_0$ \quad & \quad $w$ \quad \\ \midrule\midrule
1 & \quad 3.45 \quad & \quad 0.59 \quad \\ \midrule
0.28 & \quad 4.00 \quad & \quad 0.51 \quad \\ \bottomrule
\end{tabular}
\caption{Parameters for the empirical fit with Eq.~\eqref{eq:fit} for the flux amplification factor for an equal-mass binary NS and a NS--BH binary with $q=0.28$.
\label{tab:params}}
\end{table}

\section{Dark matter capture rate in binaries}

DM particles are captured by the NS if their orbit intersects the star and the particles lose enough energy through collisions with other particles in the NS to become gravitationally bound. In case of a binary NS, the second condition translates to the particle losing enough energy to remain within the Hill sphere of the star. As the escape velocity on the surface of the NS is much higher than the ambient DM velocity, the DM particles will almost certainly lose more energy than their kinetic energy at infinity during a single scattering, as long as their mass is not significantly higher than the nucleons they scatter off. For isolated NSs and a nucleon mass of $1$~GeV this approximation is valid for DM particles with masses $m_\chi\lesssim 10^6$~GeV. For a DM particle to remain within the Hill sphere of the NS in a binary it needs to get rid of an extra binding energy:
\begin{align}
    \Delta E_b &\approx \frac{Gm_2}{r_H} = \frac{GM}{a} \left(\frac{3m_2^2}{M^2}\right)^{1/3} \nonumber \\
    &= \frac{V_b^2}{2} 2\sqrt[3]{3}\left( \frac{q}{1+q} \right)^{2/3} \: ,
\end{align}
where we assumed that the NS has a smaller or equal mass $m_2\leq m_1$ compared to the other object. This implies that the extra binding energy only becomes comparable to the kinetic energy of the ambient DM particles when $v_\chi/V_b\lesssim1$. The most stringent constraint on DM interaction rates can be put based on binaries that merge approximately in a Hubble time. For such binary NSs and assuming a DM velocity dispersion $v_\chi=220$~km/s the binary spends most of its lifetime $v_\chi/V_b\gtrsim1/\sqrt{10}$. Since the maximum mass for which the single-scattering assumption is valid scales with $m_\chi^{\mathrm{max}}\sim (E^\chi_\mathrm{kin}+\Delta E_b)^{-1}$ \cite{Press1985}, this means the assumption remains valid in binaries for DM particles with masses up to $m_\chi\lesssim10^5$~GeV.

For particles crossing the surface of NSs relativistic effects also become relevant. This modifies the cross section, which then for $v \ll v_\mathrm{esc}$ become
\begin{align}
    \Sigma_s(v) &\approx \frac{2Gm}{v^2} \pi r_\mathrm{NS} \left( 1-\frac{2Gm}{r_\mathrm{NS}c^2} \right)^{-1} \nonumber \\
    &= \pi r_\mathrm{NS}^2 \frac{v_\mathrm{esc}^2}{v^2}\left( 1-\frac{v_\mathrm{esc}^2}{c^2} \right)^{-1} \: ,
\end{align}
or for a Maxwell-Boltzmann velocity distribution
\begin{equation}
    \left\langle\Sigma_s\right\rangle \approx \sqrt{6\pi} r_\mathrm{NS}^2 \frac{v_\mathrm{esc}^2}{v_\chi^2}\left( 1-\frac{v_\mathrm{esc}^2}{c^2} \right)^{-1} \: .
\end{equation}

So far we have only considered the geometrical cross section and flux of DM particles accreted on NSs, which accurately represents the total capture rate as long as the microscopic cross section $\sigma$ is larger than some threshold $\sigma_\mathrm{th}$ for which each DM particle is expected to scatter on a nucleon at least once while traversing through the NS. For a lower cross section, only some fraction of the incoming DM particles will go through scatterings. Then the total capture rate can be approximated by \cite[e.g.][]{Singh:2022wvw}
\begin{equation}
    \mathcal{F}^{\mathrm{capt}} \approx \mathcal{F} \cdot \min\left[ \frac{\sigma}{\sigma_\mathrm{th}}, 1 \right] \: ,
\end{equation}
where
\begin{equation}
    \sigma_\mathrm{th} = \frac{\pi r_\mathrm{NS}^2}{N_b \xi} \: .
\end{equation}
Here $N_b$ is the total number of baryons in the NS and
\begin{equation}
    \xi = \frac{4\pi}{N_b} \int\limits_0^{r_\mathrm{NS}} \mathrm{d}r \, r^2 \sqrt{g_{rr}(r)} n_b(r) \frac{1-g_{tt}(r)}{1-g_{tt}(r_\mathrm{NS})} \frac{g_{tt}(r_\mathrm{NS})}{g_{tt}(r)} \: ,
\end{equation}
where $g_{rr}$ and $g_{tt}$ are the spatial and temporal components of the Tolman--Oppenheimer--Volkoff metric inside the NS and $n_b$ is the baryon number density. Note that $\sigma/\sigma_\mathrm{th}$ is related to the optical thickness of the NS and $\xi$ contains information about the inner structure of the NS. While $\sigma_\mathrm{th}$ thus depends on the structure and equation of state of the star, for a typical NS with $m_\mathrm{NS}=1.4~M_\odot$, radius $r_\mathrm{NS}\approx 12$~km and baryon number $N_b\approx2\cdot10^{57}$, we can assume $\xi\approx1$, which makes the cross section threshold $\sigma_\mathrm{th}\approx2\cdot10^{-45}$~cm$^2$ \cite{Singh:2022wvw}. Then the total capture rate becomes
\begin{align}
    \left\langle\mathcal{F}_b^{\mathrm{capt}}\right\rangle \approx& \sqrt{6\pi} r_\mathrm{NS}^2 n_\chi \frac{v_\mathrm{esc}^2}{v_\chi}\left( 1-\frac{v_\mathrm{esc}^2}{c^2} \right)^{-1} \nonumber \\
    &\times\left\langle R_q\right\rangle \min\left[ \frac{\sigma}{\sigma_\mathrm{th}}, 1 \right] \: .
    \label{eq:DMflux}
\end{align}
The total number of DM particles accreted until the binary merges will be:
\begin{equation}
    N_b^{\mathrm{capt}} = \int\limits^{t_\mathrm{GW}} \left\langle\mathcal{F}_b^{\mathrm{capt}}\right\rangle \, \mathrm{d}t \: .
\end{equation}
Here the GW inspiral time for a circular binary is
\begin{equation}
    t_\mathrm{GW} = \frac{5}{256} M^{-3} \frac{(1+q)^2}{q} a^4 \: .
    \label{eq:tGW}
\end{equation}
The total amplification of captured DM particles during the inspiral time will then be:
\begin{equation}
    \frac{N_b^{\mathrm{capt}}}{N_s^{\mathrm{capt}}} =\frac{1}{t_\mathrm{GW}} \int\limits^{t_\mathrm{GW}} \left\langle R_q \right\rangle \, \mathrm{d}t \: .
\end{equation}
\begin{figure}[!t]
    \centering
    \includegraphics[width=0.49\textwidth]{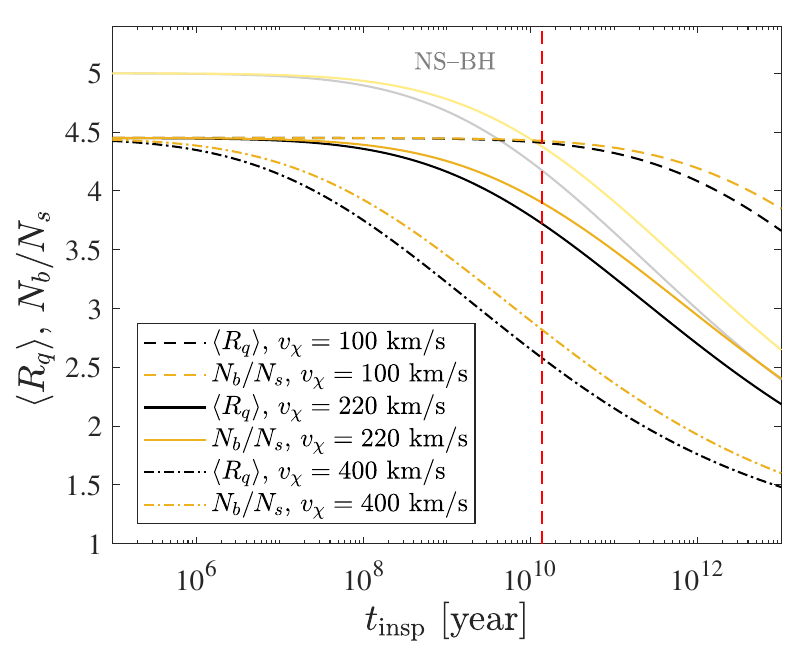}
    \caption{The averaged amplification factor $\left\langle R_q \right\rangle$ and the amplification in the total number of accreted particles $N_b/N_s$ as a function of the binary GW inspiral time. Different lines show results for different DM velocity dispersions with the solid line denoting the nominal value of $v_\chi=220$~km/s, while the dashed and dashed-dotted lines show results for high- and low-mass galaxies, respectively. The two lines with lighter tones correspond to NS--BH binaries with $q=0.28$ and $v_\chi=220$~km/s. The vertical dashed line denotes the age of the Universe.}
    \label{fig:Rtinsp}
\end{figure}
Fig.~\ref{fig:Rtinsp} shows the amplification factor as a function of the inspiral time. For the nominal DM velocity dispersion of $v_\chi=220$~km/s the amplification factor is already $\sim4$ when the binary is separated such that it would require a Hubble time to merge, assuming that only GW radiation is hardening the binary.

\section{Constraints from DM capture}

Although in general stellar objects that end up as NSs in a binary NS merger can spend different amounts of time in the binary NS phase, it is expected that their lifetime is dominated by the time spent in the GW-driven binary NS phase \cite{PortegiesZwart:1997ugk,Belczynski:2001uc,Belczynski:2006br,2019MNRAS.490.3740N}. Therefore, considering the DM capture history of NS that produce a merger, the above amplification factor cannot be neglected. After looking at the timescale of some processes that could disrupt binaries or make the inspiral time shorter, in this section we introduce two types of constraints related to DM capture in binary NSs. Since the amplification factor for NS--BH binaries does not differ significantly from that of binary NSs, here we only consider the binary NS scenario.

\subsection{Relevant timescales}
\label{ssec:timescales}

The analysis and the constraints derived in this work rely on the assumption that the binary NS lifetime is on the order of a Hubble time. However, it is a topic of strong debate wether NS binaries --and compact binaries in general-- can be formed in isolation or dynamically due to the influence of their environment \citep[see e.g][]{2008Ivanova,2019Andrews}. Therefore, here we compile a short list of physical processes that can drive binary evolution and evaluate their characteristic timescales with respect to isolated binary evolution.

We start by establishing a reference given by the GW inspiral timescale. For an equal mass binary on a circular orbit it reads \citep{1963peters}:
\begin{align}
    t_{\rm insp} \approx 16.3 \, {\rm Gyr}\, \times \left( \frac{a}{0.03\, {\rm AU}}\right)^4\left(\frac{4\,{\rm M}_{\odot}}{M} \right)^{3},
\end{align}
meaning that even very compact binaries with $a \lesssim 0.1$ AU will inspiral for at least a cosmic time.

Binary evolution can be driven by perturbations of a tertiary component by the von Zeipel-Kozai-Lidov effect \cite{zeipel1910,koz62,lid62}, which induces eccentricity oscillations that can result in more efficient merger \citep{2013seto,2019liu}. The typical timescale of ZKL oscillations is given by the ratio of the outer period $P_{\rm out}$ squared to the inner binary period $P_{\rm in}$:
\begin{align}
    t_{\rm KZL} \approx \frac{M}{M_3} \frac{P_{\rm out}^2}{P_{\rm in}}(1-e_{3}^2)^{3/2},
\end{align}
where $M_3$ and $e_3$ are the mass and the eccentricity associated to the ``outer" orbit of the perturber. For a NS binary of 4 M$_{\odot}$ at 0.03 AU, the orbital period $P_{\rm in}\sim$1 day. Then, we can estimate the KZL timescale:
\begin{align}
    t_{\rm KZL}^{\rm star} &\approx \frac{4\, \rm{M}_{\odot}}{M_3} \frac{P_{\rm out}^2}{1\, {\rm day}},
\end{align}
where we assumed a circular outer orbit for simiplicity. We see how the only scenario in which ZKL oscillation can realistically influence the inspiral timescale of the binary below $\sim 0.1$ AU is in a hierarchical triple system, in which the outer period is smaller than:
\begin{align}
    P_{\rm out} < 6\times 10^3\left( \frac{M_3}{4\, \rm{M}_{\odot}}\right) \, \rm{yr},
\end{align}
where we equated the KZL timescale with a Hubble time. For equal mass components, this corresponds to an outer separation of $\sim 600$ AU.

Stellar flybys and random encounters can also influence the NS binary. The rate of flybys depends on the density $n_{\rm star}$ and the velocity dispersion $v_{\rm star}$ in the surroundings of the binary NS. The timescale for an encounter with an impact parameter comparable to the binary semi-major axis is:
\begin{align}  
 t_{\rm enc}^{\rm field} &\approx\frac{1}{n_{\rm star}v_{\rm star} a^2}.
\end{align}
In a dense stellar environment, such as a nuclear cluster, typical parameters read \citep{2016fritz,2024hannah}:
\begin{align}
    t_{\rm enc}^{\rm NC} 
    &\approx 2.3\times10^2 \, {\rm Gyr}\left(\frac{10^{6}\, {\rm pc}^{-3}}{n_{\rm star}} \right)\nonumber \\ &\times\left(\frac{200 \, {\rm km/s}}{v_{\rm star}} \right)\left(\frac{0.03\, {\rm AU}}{a} \right)^2.
\end{align}
We find that flybys and encounters will only significantly perturb the binary in extremely dense stellar system, typically denser than even nuclear clusters.

Finally, we model the effect of gas accretion if the binary is evolving in a gas rich environment. This can be the case for binaries in star forming clusters \citep{2023rozner}, or in a more extreme case binaries embedded in an AGN accretion disc \citep{2020tagawa}. We model the accretion flow with the Bondi--Hoyle--Lyttleton prescription \citep{1944bomdi}, though note that in reality accretion on NS binaries will be significantly more complex due to magnetic fields, feedback and other processes. The timescale to modify the orbit via accretion is comparable to the mass doubling timescale:
\begin{align}
    t_{\rm BHL} & \approx \frac{M}{\dot{M}_{\rm BH}},
\end{align}
where BH accretion is:
\begin{align}
    \dot{M}_{\rm BH} = \pi\frac{ G^2 M^2 }{c_{\rm s}^3}\rho_{\rm gas}.
\end{align}
Then, the timescale for BHL accretion is:
\begin{align}
    t_{\rm BHL} &\approx 4.8\times 10^3\,{\rm yr} \left(\frac{10^{-13}\,{\rm M_{\odot}/AU^3}}{\rho_{\rm gas}}\right)  \nonumber\\
    &\times \left(\frac{c_{\rm s}}{100\, {\rm m/s}}\right)^{3}\left(\frac{4\, {\rm M}_{\odot}}{M}\right) 
\end{align}
where we scaled with typical parameters for cold hydrogen in a star forming globular cluster \citep{2022lin}. This shows that gas accretion is an extreme efficient driver of binary evolution, even at the compact scales of $a\lesssim 0.1$ AU, though we note again that this is a highly idealized accretion model. While the efficiency of gas accretion would suggest that our assumption may fail, we note that accretion also drives the NSs towards their maximum stable mass and collapse into BHs well before they merge. Hence, this does not affect our constraint for the DM interaction cross section, which assumes the observation of a binary NS merger.

\subsection{Condensation and BH formation}

If the accreted DM particles do not annihilate then the capture and accumulation of these particles can lead to BH formation at the NS core and the subsequent implosion of the NS itself. In case of the simplest non-interacting DM scenario, there is a huge difference between the Chandrasekhar limit for fermionic and bosonic particles, since the former is supported by the degeneracy pressure caused by the exclusion principle, while the latter can form a dense core inside the NS due to Bose--Einstein condensation. The number of DM particles required for fermionic and bosonic particles to reach the Chandrasekhar limit is in fact \cite{Singh:2022wvw}:
\begin{align}
    N_\mathrm{lim}^\mathrm{f} &\approx 1.8 \times 10^{57} \: \left( \frac{\mathrm{GeV}}{m_\chi} \right)^{3} \: , \\
    N_\mathrm{lim}^\mathrm{b} &\approx 1.5 \times 10^{38} \: \left( \frac{\mathrm{GeV}}{m_\chi} \right)^{2} \: .
\end{align}
Notice that the total mass required for fermionic DM particles to collapse into a BH is of the order of the NS mass itself. This raises the question whether the binaries can accumulate such a large amount of DM particles without these particles also significantly accelerating the inspiral process. According to Ref.~\cite{Quinlan:1996vp} the hardening rate $\dot{a}/a$ of binaries due to dynamical friction is proportional to the flux of particles entering the sphere with radius $a$, thus:
\begin{equation}
    \frac{\dot{a}}{a} \sim \mathcal{O}(1) \frac{m_\chi}{M} n_\chi v_\chi \Sigma^{a}(v_\chi) \: ,
\end{equation}
where $\Sigma^a$ is the gravitational cross section given by Eq.~\eqref{eq:GrCrSec} at $r=a$. Meanwhile the mass growth of the NS is
\begin{equation}
    \frac{\dot{m}_\mathrm{NS}}{m_\mathrm{NS}} \sim \mathcal{O}(1) \frac{m_\chi}{M} n_\chi v_\chi \Sigma^{r_\mathrm{NS}}(v_\chi) \: .
\end{equation}
Therefore, since $\Sigma^{a}/\Sigma^{r_\mathrm{NS}}\gg 1$ the hardening rate due to dynamical friction by DM particles is many orders of magnitude higher than the mass growth of the NS. This means that the binary will inspiral much before it could double its mass. This does not pose a problem for bosonic DM, which require a total DM mass $\sim 20$ orders of magnitude lower to form a small BH inside the NS core. We will discuss the hardening of binaries due to interactions with the ambient DM in further detail in Sec.~\ref{ssec:DMfrac}.

There are two pathways for bosonic DM particles to form a BH described in detail in Ref.~\cite{Singh:2022wvw}. After the particles have been captured by the NS, they thermalize with baryonic matter on the timescale of $10^{-1}-10^3$~years. Then if no Bose--Einstein condensation occurs, a BH will form approximately when the central mass density of the DM particles exceeds the density of baryonic particles and they become self-gravitating. The number of DM particles required for this to happen is \cite{Bertoni:2013bsa,Singh:2022wvw}:
\begin{equation}
    N^\mathrm{BH}_\mathrm{self} \approx 4.8 \times 10^{46}\left( \frac{\mathrm{GeV}}{m_\chi} \right)^{5/2} \left( \frac{T}{10^5~\mathrm{K}} \right)^{3/2} \: .
\end{equation}
However, for lower mass DM particles the critical temperature for Bose--Einstein condensation of DM can surpass the NS temperature before the DM particles can become self-gravitating, at which point a concentrated region of condensed bosons appears in the center of the NS. The number of particles required for the condensation to occur is \cite{McDermott:2011jp,Kouvaris:2011fi,Singh:2022wvw}:
\begin{equation}
    N_\mathrm{BEC} \approx 10^{36}\left( \frac{T}{10^5~\mathrm{K}} \right)^{3} \: .
\end{equation}
Thus, for an old NS with $T=10^5$~K this happens earlier for $m_\chi\lesssim2\cdot 10^4$~GeV, while for $T=10^6$~K the same limit is $m_\chi\lesssim5\cdot 10^3$~GeV. After the condensate has been formed, it should also reach the Chandrasekhar limit, which requires $N_\mathrm{lim}^\mathrm{b}$ additional particles. This makes the total number of DM particles needed for BH formation through Bose--Einstein condensation:
\begin{equation}
    N^\mathrm{BH}_\mathrm{BEC} \approx 10^{36} \left[ \left( \frac{T}{10^5~\mathrm{K}} \right)^{3} + 1.5 \times \left( \frac{10\,\mathrm{GeV}}{m_\chi} \right)^{2}  \right] \: .
\end{equation}

After the formation of a small BH inside the NS core, it will start devouring the matter of the NS, which can happen if the accretion rate is higher than the mass loss due to Hawking radiation. For low $m_\chi$ the BH evaporation is suppressed by the accretion of NS matter, while for higher DM masses it can be suppressed by the accretion of DM itself \cite{McDermott:2011jp,Garani:2018kkd,Bell:2020jou,Bertoni:2013bsa,Kouvaris:2011fi,Singh:2022wvw}. Overall, an unstable BH is formed in the Bose--Einstein condensation scenario, if $m_\chi\gtrsim10$~GeV \cite{Khadkikar:2025awf}. In every other case the BH is stabilized by the accretion of matter. Then, the small BH will start devouring the NS, which, according to Refs.~\cite{Baumgarte:2021thx,Singh:2022wvw}, happens on the Bondi-Hoyle accretion timescale of $10^4-10^7$~years, depending on the mass of the DM particle.

\subsection{Constraint on bosonic DM cross section}

Based on the above, a central BH is formed if the number of accreted DM particles reaches the value:
\begin{equation}
    N^\mathrm{BH} = \min\left( N^\mathrm{BH}_\mathrm{self},N^\mathrm{BH}_\mathrm{BEC} \right) \: .
\end{equation}
Assuming that the binary had approximately a Hubble time, $t_H$, to accumulate this amount of DM, the thermalization and implosion timescales can be neglected. Then based on the observation merging binary NSs (that have not collapsed into BHs) in the local Universe the following constraint can be put on the scattering cross section of baryonic and DM particles:
\begin{equation}
    \sigma \lesssim \frac{N^\mathrm{BH}}{N_b(t_H)} \sigma_\mathrm{th} \: ,
\end{equation}
where $N_b(t)$ is the number of particles accreted during the inspiral time $t$ when the microscopic cross section is high, i.e. $\sigma>\sigma_\mathrm{th}$:
\begin{equation}
    N_b(t) = \int\limits^{t} \left\langle\mathcal{F}_b\right\rangle \, \mathrm{d}t'
\end{equation}
The above constraint is only valid for DM masses where the upper bound on $\sigma$ is below $\sigma_\mathrm{th}$.
\begin{figure}[!t]
    \centering
    \includegraphics[width=0.49\textwidth]{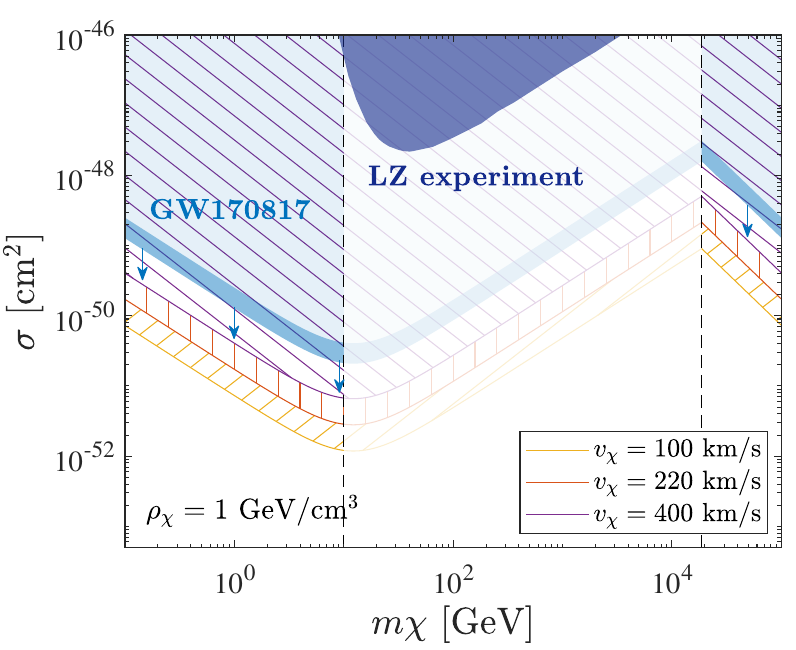}
    \caption{Upper bounds on the microscopic scattering cross section of between baryonic and DM particles based on the observation of isolated binary NS mergers embedded in a DM environment of with density $\rho_\chi=1$~GeV/cm$^3$ and various values for the DM velocity dispersion $v_\chi$. The vertical dashed line at $m_\chi\approx2\times10^4$~GeV is denoting the limit below which the central BH is formed through Bose--Einstein condensation, while above this limit no Bose--Einstein condensation takes place. The range of upper bounds from GW170817 (blue filled region) is calculated with $\rho_\chi=0.1$~GeV/cm$^3$, $v_\chi=220$~km/s, and the 90\% confidence range of delay times $(6.8-13.6)$~Gyr inferred by Ref.~\cite{Blanchard:2017csd}. The constraints from DM accretion on NS do not apply in the region between the vertical dashed lines, where the BH seeds are unstable against Hawking radiation. The purple filled region at the top shows constraints from the LZ experiment \cite{LZ:2022lsv,LZ:2024zvo}.}
    \label{fig:CrossSec}
\end{figure}
Fig.~\ref{fig:CrossSec} shows our results for the upper bounds on the DM interaction cross section with nucleons for $t_H\approx10^{10}$~yr and old NSs with $T\approx10^5$~K in the relevant mass range of $0.1~\mathrm{GeV} \leq m_\chi \leq 10^5~\mathrm{GeV}$. For low DM masses Bose--Einstein condensation can set in before the DM component becomes self-gravitating, which then reduces the number of particles required for BH formation in that mass range. However, a BH stable against Hawking radiation can only form in this case if $m_\chi\lesssim10$~GeV \cite{Khadkikar:2025awf}. Lower velocity dispersions provide higher accretion rates both due to the increased gravitational cross section $\Sigma_s$ and the increased binary amplification rate. Fig.~\ref{fig:CrossSec} also shows upper bounds from the direct detection LUX-ZEPLIN (LZ) DM experiment \cite{LZ:2022lsv,LZ:2024zvo}. Note how due to instability of BHs in the Bose--Einstein condensation scenario above $\sim10$~GeV, NSs provide constraints complementary to the LZ experiment.

\subsubsection{Constraint from GW170817}

The GW signal GW170817 was unambiguously associated with a binary NS merger \cite{LIGOScientific:2017ync}. Follow-up studies of the host galaxy NGC 4993 inferred that the star formation inside the galaxy peaked at $\gtrsim10$~Gyr ago and subsequently showed the GW delay time of the progenitor of GW170817 to be $(6.8-13.6)$~Gyr \cite{Blanchard:2017csd}. Other works have also analyzed the DM profile of the galaxy derived from stellar kinematics \cite[e.g.][]{Ebrova:2018gtz,Gaspari:2024abt} and have inferred a central DM density of $\sim7\times10^{-3}~M_\odot/\mathrm{pc}^3\approx0.27$~GeV/cm$^3$ \cite{Gaspari:2024abt}. Even though the estimated scale radius of $\sim46^{+22}_{-15}$~kpc \cite{Gaspari:2024abt} of the DM density profile is much larger than the inferred distance, $\sim2$~kpc, of the merger event from the galactic center, we use a conservative value of $0.1$~GeV/cm$^3$ for the DM density. Additionally, we use $v_\chi=220$~km/s as a conservative value for the DM velocity dispersion.

The constraints on the bosonic DM cross section derived from this observation are also shown in Fig.~\ref{fig:CrossSec}.

\subsection{Maximum DM fraction at merger}
\label{ssec:DMfrac}

As the NSs inside the binary accumulate DM, interactions with the surrounding particles will also necessarily extract energy from the binary. This means that the total amount of DM accumulated in these NSs will be limited by the fact that the surrounding DM also drives the binary towards merger. The total hardening rate of the binary will be the sum of the hardening due to dynamical friction and that due to GW radiation:
\begin{equation}
    \dot{a}_\mathrm{tot} = \dot{a}_\chi + \dot{a}_\mathrm{GW} \: ,
\end{equation}
where
\begin{equation}
    \dot{a}_\chi = -a^2 \frac{G\rho_\chi}{v_\chi} H(v_\chi) \: ,
\end{equation}
with $H(v_\chi)$ being a numerical factor, which depends on the DM velocity dispersion \cite{Quinlan:1996vp}. Since for typical DM velocities considered in this paper $v_\mathbf{DM}/V_b\lesssim0.5$, we can assume that for the the circular and equal-mass binaries considered here, $H=\mathrm{const.}\approx28$. The hardening rate due to GW emission is:
\begin{equation}
    \dot{a}_\mathrm{GW} = -\frac{64G^3}{5c^5} M^3 \frac{q}{(1+q)^2}a^{-3}.
\end{equation}
Due to the large difference in the power-law dependence of the hardening rates on $a$, the evolution of the binary will be separated into two regimes. For large separations, drag forces will be dominated by dynamical friction by DM, with GW radiation drag being negligible, while for small separations GW radiation will dominate over dynamical friction. The separation at which the two effects are equal is given by:
\begin{equation}
    a^* \approx \left( \frac{16G^2}{35c^5} M^3 \frac{q}{(1+q)^2} \frac{v_\chi}{\rho_\chi} \right)^{1/5} \: ,
    \label{eq:astar}
\end{equation}
where we used $H\approx28$.

The DM drag dominated phase lasts for:
\begin{equation}
    t_\mathrm{DM} = \frac{v_\chi}{G\rho_\chi H}\left( \frac{1}{a^*} - \frac{1}{a_0} \right) \: ,
\end{equation}
where $a_0$ is the initial semi-major axis of the binary. This means that the time it takes for the binary to get through this phase is at most:
\begin{equation}
    t_\mathrm{DM}^\mathrm{max} = \frac{v_\chi}{G\rho_\chi H} a^{*-1} \: .
\end{equation}
The GW inspiral time is given by Eq.~\eqref{eq:tGW}:
\begin{equation}
    t_\mathrm{GW} = \frac{5c^5}{256G^3} M^{-3} \frac{(1+q)^2}{q} a^{*4} \: .
\end{equation}
Assuming that the binary merges within a Hubble-time, we need to have $t_\mathrm{H}\leq t_\mathrm{DM} + t_\mathrm{GW}$.

Defining the accreted DM mass fraction as $f_\chi \equiv \Delta m_\chi/m_\mathrm{NS}$, and assuming that $\sigma\geq \sigma_\mathrm{th}$, i.e. the DM capture rate is maximal and given by the geometrical flux, and $\Delta m_\chi \ll m_\mathrm{NS}$ we have from Eq.~\eqref{eq:DMflux}:
\begin{align}
    \dot{f}_\chi &= \sqrt{6\pi} \rho_\chi r_\mathrm{NS} \frac{2G}{v_\chi} \left\langle R_q\right\rangle \left( 1-\frac{v_\mathrm{esc}^2}{c^2} \right)^{-1} \nonumber \\
    &\approx 14 \sqrt{6\pi} \frac{G\rho_\chi r_\mathrm{NS}}{v_\chi} \: ,
\end{align}
where we have used $\left\langle R_q\right\rangle\approx4.5$. Using this we can now put an upper bound on the total DM fraction of binary NSs at merger given that they spent most of their lifetime in a binary system.

Two scenarios can be separated. If, for a given DM density $\rho_\chi$, $t_\mathrm{DM}^\mathrm{max} + t_\mathrm{GW} \geq t_\mathrm{H}$, then the binary can accrete for $t_\mathrm{H}$ and the total accumulated DM fraction becomes:
\begin{equation}
    f^b_\chi \approx 14 \sqrt{6\pi} \frac{G\rho_\chi r_\mathrm{NS}}{v_\chi} t_\mathrm{H} \: .
\end{equation}
However, if $\rho_\chi$ is such that $t_\mathrm{DM}^\mathrm{max} + t_\mathrm{GW} < t_\mathrm{H}$ then the binary will merge within a Hubble time. Then the accumulated DM fraction in the DM dominated phase becomes:
\begin{equation}
    f_\mathrm{DM} \approx \frac{\sqrt{6\pi}}{2} \frac{r_\mathrm{NS}}{a^*} \: ,
\end{equation}
while in the GW dominated phase:
\begin{equation}
    f_\mathrm{GW} \approx 14 \sqrt{6\pi} \frac{G\rho_\chi r_\mathrm{NS}}{v_\chi} t_\mathrm{GW} \: .
\end{equation}
Then the total DM fraction is $f_\chi^b = f_\mathrm{DM} + f_\mathrm{GW}$. On the other hand, the DM fraction accumulated in a single NS over a Hubble time is:
\begin{equation}
    f_\chi^s \approx 3 \sqrt{6\pi} \frac{G\rho_\chi r_\mathrm{NS}}{v_\chi} t_\mathrm{H} \: .
\end{equation}

\begin{figure}[!t]
    \centering
    \includegraphics[width=0.49\textwidth]{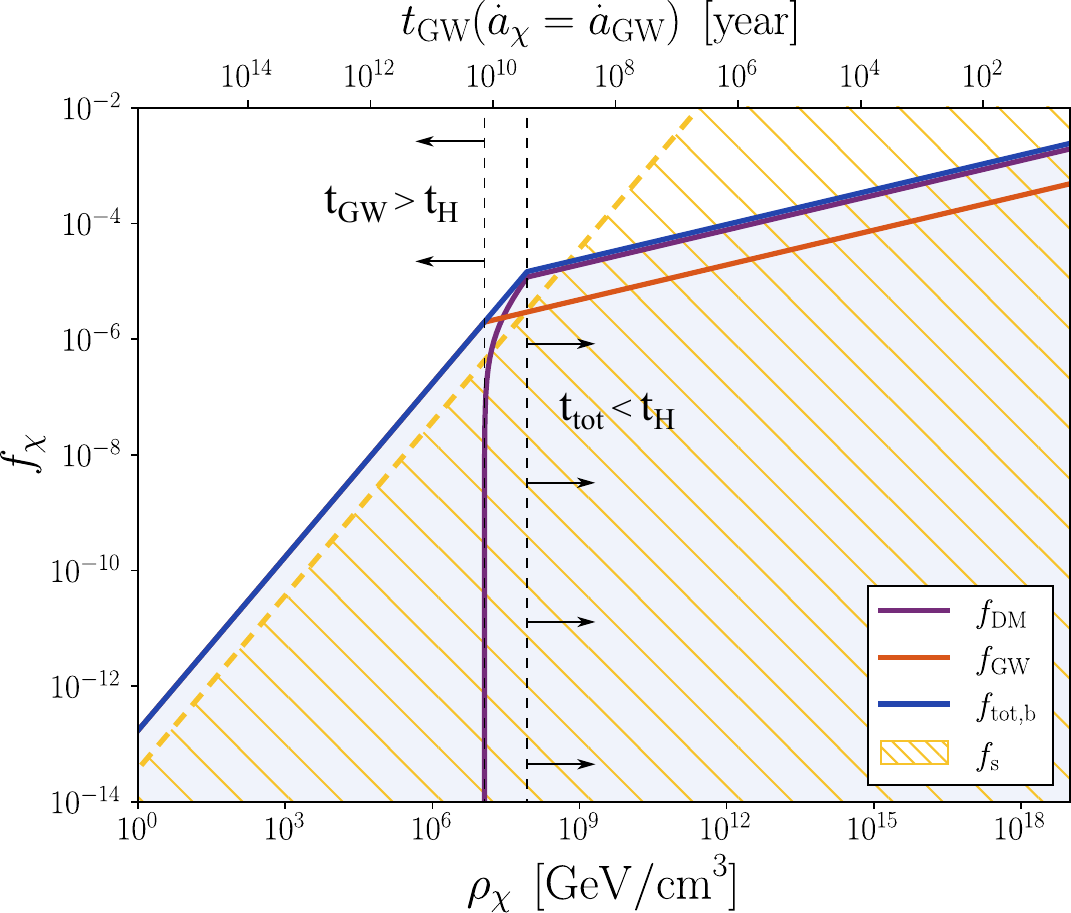}
    \caption{Upper bound on the mass fraction of accreted DM assuming that the NSs spend most of their lifetime in a binary system, embedded in a homogeneous DM medium with density $\rho_\chi$ and velocity dispersion $v_\chi=220$~km/s (blue). For low densities, DM is accumulated during the GW radiation dominated phase (red), while for high densities, most of the DM is accumulated during the DM drag dominated phase (purple). The upper bound on DM mass fraction for single NSs is shown by the yellow hatched region. For low DM densities, this is below that for binary systems, while for large densities, it continues to scale with $f_\chi\propto\rho_\chi$ and thus overshoots the constraint for binary systems.}
    \label{fig:DM_frac}
\end{figure}
Fig.~\ref{fig:DM_frac} shows the constraints on the maximum possible DM fraction accumulated in binaries by the time of merger given that they are embedded in a homogeneous medium of DM with a density $\rho_\chi$. Here we used $v_\chi = 220$~km/s. For low DM densities, $t_\mathrm{GW}>t_\mathrm{H}$ and the binary dynamics is always dominated GW radiation. Then the total DM fraction simply scales with $f_\chi\propto\rho_\chi$. For densities where $t_\mathrm{GW}<t_\mathrm{H}$ but $t_\mathrm{DM}^\mathrm{max}+t_\mathrm{GW}>t_\mathrm{H}$ this scaling continues but some fraction of the DM will be accumulated during the DM drag dominated phase. However, for high densities with $t_\mathrm{DM}^\mathrm{max}+t_\mathrm{GW}<t_\mathrm{H}$ accreted DM is given mostly by the DM dominated phase and the scaling becomes $f_\chi\propto\rho_\chi^{1/5}$. This significantly limits the maximum amount of DM that can be accreted. Note, however, that if the NSs spend most of their lifetime as singles and only form a binary later, then this constraint does not apply and the total DM fraction continues to scale as $f_\chi\propto\rho_\chi$ only with a smaller coefficient.

We note that environmental effects, such as DM drag in this case, if strong enough, can be inferred from the dephasing of the binary GW signal \cite[e.g.][]{Samsing:2024syt,Hendriks:2024gpp,Zwick:2025wkt,Takatsy:2025bfk}. Dephasing due to interaction with the surrounding DM has been the topic of multiple recent studies \cite[e.g.][]{Cardoso:2019rou,Speeney:2022ryg,Wilcox:2024sqs}. In our case the GW frequency corresponding to $a^*$ in Eq.~\eqref{eq:astar} scales as:
\begin{align}
    f^*_\mathrm{GW} \approx 2.4\times 10^{-4}&\left( \frac{\rho_\chi}{10^{-10}~\mathrm{GeV/cm}^3} \right)^{3/10} \nonumber \\ 
    &\times\left( \frac{M}{3~M_\odot} \right)^{-2/5} ~\mathrm{Hz} \: .
\end{align}
Thus, based on this the dephasing due to DM drag might be detectable with the LISA GW detector \cite{LISA} for the highest DM densities of $\rho_\chi\gtrsim10^{10}$~GeV/cm$^3$. We also note that the dephasing can already become detectable for inspirals with large signal-to-noise ratios if the DM density is 1-2 orders of magnitude lower than this.

\section{Discussion}
\label{sec:conclusion}

In this work we showed how the flux of non-interacting particles on a central gravitating object can be amplified in binary systems. Using Monte Carlo simulations and the scaling property of the problem we calculated this amplification factor as function of the dimensionless parameter $v/V_b$, for two different mass ratios and vanishing eccentricities. We also provided an empirical fit for the results. Using these fits we determined how much the total number of accreted DM particles is amplified in a binary system compared to isolated objects, for realistic DM velocity dispersions.

Based on this amplification we provided improved upper bounds on the microscopic scattering cross section of DM and baryonic particles, compared to \cite{Singh:2022wvw,Khadkikar:2025awf}. This constraint is based on the hypothetical observation of NS binary mergers in environments with dense DM halos. We also derived a more robust constraint based on the observation of GW170817 and its electromagnetic counterparts, as well as the inferred delay time of the merger and properties of the host galaxy. Given the anticipated rise of the number of binary mergers with masses corresponding to NSs, observed by LVK in the near future, it would become possible to provide upper and lower bounds as well on the cross section. This could be based on the idea of estimating the collapse time of NSs due to DM accretion inferred from the observed fraction of collapsed NSs in such mergers (i.e. the lack of observation of electromagnetic signals or signs of tidal deformability from these mergers), as suggested by Ref.~\cite{Singh:2022wvw}, and recently showed by Ref.~\cite{Khadkikar:2025awf} through GW190425. We note, however, that as discussed in Sec.~\ref{ssec:timescales}, gas accretion can also lead to the collapse of the NS into a BH, although such a process would only produce BHs above the -- yet uncertain -- TOV limit of NSs and a BH with a mass of $\sim1.4~M_\odot$ could not be the result of gas accretion. Tidal deformability measurements with the LVK detectors is currently also not accurate enough to tell these different scenarios apart confidently. While this accuracy is expected to improve in the future, the merger event GW170817 already provides a robust upper bound on the DM cross section.

Finally we also provided constraints for the maximum possible amount of DM that can be accreted on NSs during their binary NS phase. These constraints do not only apply to the bosonic DM candidates but fermionic ones as well. Based on these calculations only a very small fraction of the NS mass can be accumulated in DM for realistic DM densities. While it is hypothesized that in the very center of galaxies, inside DM spikes, the local DM density can reach values of $(10^{10}-10^{18})$~GeV/cm$^3$ \citep[e.g.][]{Sadeghian:2013laa}, in such dense environments other stellar objects would also be able to disrupt the binary much more efficiently. We note that the accumulated DM is not limited by dynamical friction in these dense environments if the NS spend most of its lifetime as an isolated object. However, according to current predictions on the possible formation channels of binary NS mergers, these binaries most likely form through the isolated binary channel \cite{PortegiesZwart:1997ugk,Belczynski:2001uc,Belczynski:2006br,2019MNRAS.490.3740N} and therefore it seems unlikely that a large number of binary NS mergers are produced from NSs that spend a significant amount of time in isolation. Moreover, due to mass segregation \cite{1969ApJ...158L.139S}, these NSs are expected to be kicked out of the galactic center on a timescale that is a small fraction of a Hubble time. Finally, we also note that for larger scattering cross sections the NSs could accumulate more matter during their stellar phase, given their size and thus a larger geometric flux of DM particles. However, such scattering cross sections become increasingly excluded by other DM constraints.

As a final remark, we note that our calculations only apply for binaries with a negligible eccentricity. For finite eccentricities, the amplification factor, the effect of dynamical friction, as well as the GW inspiral timescale would change as well. Even though for a fixed semi-major axis we do not expect a significant change in the amplification factor, and indeed, as shown by Ref.~\cite{Quinlan:1996vp}, dynamical friction forces only receive a correction of $\mathcal{O}(1)$, the GW inspiral time can change significantly. This means that binaries that merge with observable eccentricities spend more time at larger separations, where the amplification factor is smaller and the binary is better exposed to external perturbations that could disrupt the binary, leading to a shorter binary lifetime.

\begin{acknowledgments}
J.T. acknowledges support from the Horizon Europe research and innovation programs under the Marie Sk\l{}odowska-Curie grant agreement no. 101203883. D.ON acknowledges support from the Danish Independent Research Fund through Sapere Aude Starting Grant No. 121587. This work was supported by the ERC Starting Grant No. 121817–BlackHoleMergs led by Johan Samsing, and by the Villum Fonden grant No. 29466, which LZ acknowledges. The Center of Gravity is a Center of Excellence funded by the Danish National Research Foundation under grant No. 184. J.T. would like to thank Daniel J. D'Orazio for illuminating discussions.

\end{acknowledgments}


\bibliography{DMaccretion}


\end{document}